\documentclass[twocolumn,showpacs,amsmath,amssymb,floatfix,aps,prl]{revtex4}

\usepackage{graphicx}
\usepackage{color}

\begin{document}

\title{Effective theory of magnetization plateaux in the Shastry-Sutherland lattice}
\author{A. Abendschein$^{1,2}$}
\author{S. Capponi$^{1,2}$} 
\email{capponi@irsamc.ups-tlse.fr} 
\affiliation{{$^1$Universit\'e de Toulouse; UPS; Laboratoire de
Physique Th\'eorique (IRSAMC); F-31062 Toulouse, France}\\
{$^2$CNRS; LPT (IRSAMC); F-31062 Toulouse, France}}

\begin{abstract}
We use the non-perturbative Contractor-Renormalization method (CORE) in order to derive an effective model for triplet
excitations on the Shastry-Sutherland lattice. For strong enough magnetic fields, various magnetization plateaux are
observed, e.g. at 1/8, 1/4, 1/3 of the saturation, as found experimentally in a related compound. Moreover, 
other stable plateaux are found at 1/9, 1/6 or 2/9.  We give a
critical review of previous works and try to resolve some apparent inconsistencies between various theoretical 
approaches. 
\end{abstract}

\date{\today}

\maketitle

\begin{figure}
\begin{center}
\includegraphics[width=.38\textwidth,clip]{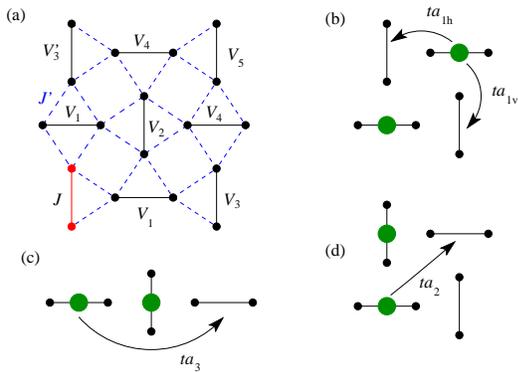}
\end{center}
\caption[]{(Color online) Shastry-Sutherland lattice and definitions of some effective interactions: (a)
  2-body interaction between a particle on the red dimer and one on the
  labelled dimers. (b) - (d) Typical correlated hopping processes. 
\label{sh_sl_int_hop.fig} }
\end{figure}

The Shastry-Sutherland lattice~\cite{Shastry1981} and
its realization in the material SrCu$_2$(BO$_3$)$_2$ have been
attracting a lot of attention due to its fascinating behaviour in a
magnetic field~\cite{Kageyama1999b,Onizuka2000,Kodama2002}, namely
that magnetization plateaux have experimentally been observed for
values of $m=1/8$, $1/4$, and $1/3$ of the
saturation value. The Shastry-Sutherland lattice, sketched in
Fig.~\ref{sh_sl_int_hop.fig}(a), is a two-dimensional (2d) Heisenberg
antiferromagnetic spin-1/2 coupled-dimer system which Hamiltonian
reads:
\begin{equation}
H = J \sum_{<i,j>} {\bf S}_i \cdot {\bf S}_j
+ J' \sum_{\ll i,j \gg} {\bf S}_i \cdot {\bf S}_j - h \sum_i S_i^z.
\label{H_gen_Sh_Sl}
\end{equation}

Experiments with SrCu$_2$(BO$_3$)$_2$ indicate that the ratio between
inter- and intra-dimer coupling should be close to $J'/J\approx 0.65$
where the ground-state is exactly given by the product of singlets on
$J$ bonds~\cite{Miyahara2003a}. In the presence of a finite magnetic field $h$, polarized
triplets are created on the dimer bonds so that 
low-energy properties can be described with an effective model of
these ``particles'': since these triplets are hard-core bosons moving
on an effective square lattice, they can typically exhibit
compressible superfluid or incompressible Mott phases depending
on the filling~\cite{Momoi2000, Miyahara2003a}. In the original language, these two phases
correspond respectively to absence or presence of magnetization
plateaux.

In order to provide a simple picture, let us recall that the triplet
hopping is strongly reduced on frustrated lattices, and therefore the
physics is governed by effective Coulomb repulsion, resulting in
various insulating phases known as Wigner crystals. Following these
pioneering works, other theoretical approaches have confirmed the
occurrence of several plateaux~\cite{Misguich2001,Miyahara2003a}. While
all approaches agree to describe $m=1/2$ or $1/3$ plateaux, the
situation is less clear at lower magnetization.

Experimentally, because of accessible fields, the first plateaux were
discovered at 1/8 and 1/4~\cite{Onizuka2000}, but recently, translation symmetry breaking has been observed above 
1/8~\cite{Takigawa2007,Levy2008} as well as evidence of 1/6 plateau~\cite{Takigawa2008}. By using torque
measurements, Sebastian {\it et al.}~\cite{Sebastian2007} have
suggested additional magnetization plateaux for more ``exotic'' values
like $1/9$, $1/7$, $1/6$, $1/5$ and $2/9$. Although an agreement on
these values has not been reached yet, these results are quite
exciting and ask for a thorough theoretical analysis.

Because of the lack of powerful numerical techniques or unbiased
analytical tools to tackle 2d frustrated systems, a promising approach
consists in deriving an effective hard-core bosonic model. Because one
is interested in low-magnetization (i.e. low filling), long-range
effective interactions are crucially needed and can only be captured
thanks to efficient algorithms. Recently, perturbative continuous unitary transformations up to high orders 
have provided such an effective model~\cite{Dorier2008}. 
In this letter, we use the \emph{non-perturbative} Contractor
Renormalization (CORE)
technique~\cite{Morningstar1994,Capponi2004} in order
to derive an effective model for the polarized triplets. Then, in
order to provide an unbiased analysis of this model, we solve it
\emph{exactly} on various clusters and by finite-size scaling
analysis, we predict the existence or not of some plateaux in the
thermodynamical limit.

{\it Effective model --} Similarly to perturbation theory~\cite{Momoi2000,Miyahara2003a}, 
we base our approach on keeping only the singlet and polarized triplet states on each dimer~\cite{Abendschein2007}.
A crucial aspect of the CORE technique is that it gives a cluster
expansion of the effective Hamiltonian ${\cal H}_{\rm eff}$. Basically, the amplitude of local processes can already be
captured by solving a small finite cluster. As a consequence, the only
approximation consists in truncating beyond a certain range of
interactions~\cite{Morningstar1994}. In our study, we keep \emph{all} processes that can
appear on a $3\times 3$ cluster (corresponding to 18 original sites). Note that 
some elementary processes, as nearest-neighbor repulsion or chemical potential, are already well 
captured with a smaller  $2\times 2$ cluster, so that different truncations may give similar amplitudes. 
Nevertheless, long-range interactions are crucially needed to describe low-filling phases. 
 Typically, ${\cal H}_{\rm eff}$ contains of the order of $10^4$
terms, similar to what is obtained with high-order
perturbation~\cite{Dorier2008}.

\begin{figure}
\begin{center}
  \includegraphics[width=.4\textwidth,clip]{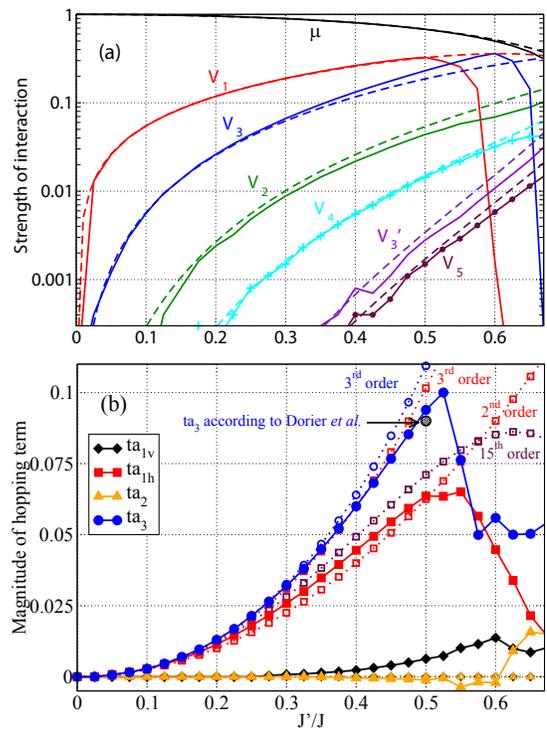}
\end{center}
\caption[]{(Color online) Effective amplitudes, defined in
  Fig.~\ref{sh_sl_int_hop.fig}, calculated with CORE (solid lines)
  and perturbation theory (dashed lines, see text): (a) $\mu$ and
  2-body diagonal terms; (b) correlated hoppings and high-order
  perturbation result for $ta_{1h}$ and $ta_3$ from Dorier {\it et al.} (private
  communication).
\label{sh_sl_interactions.fig} }
\end{figure}

Our CORE calculation gives access to the chemical potential $\mu$ which corresponds to the spin gap.
As sketched in Fig.~\ref{sh_sl_interactions.fig}(a),  CORE
results are in good agreement with 4th order perturbation theory~\cite{Miyahara2003a} up to $J'/J=0.7$. 
Because of the important role played by diagonal 2-body interactions, we compare them with 
high-order perturbation theory~\cite{Dorier2008} in Fig.~\ref{sh_sl_interactions.fig}(a).
  Clearly, these interactions
decrease with distance and are non-isotropic: for instance,  $V_3$ and $V_3'$ strongly differ. 
Generally, one also notices very good agreement between 
perturbation theory (third order~\cite{Momoi2000, Miyahara2003a} or higher~\cite{Dorier2008})
 and CORE results at least up to values of $J'/J\approx
0.5$. Beyond this value, other processes become dominant so that comparison becomes more difficult to perform and we will restrict to $J'/J\leq 0.5$ in our study. 

Besides the discussed 2-body interaction terms, $V_3^{3p}$, that corresponds to an attractive interaction of three aligned particles,  has a sizeable contribution (-0.7) close to $J'/J\sim 0.5$, quite different from high-order perturbation by Dorier {\it et al.} (-0.4). 
Although being high-order processes in perturbation,  we believe that  interaction
terms involving three and more particles play an important role in the
formation of plateaux.
 For example, in the one-dimensional realisation of the
Shastry-Sutherland lattice, namely the orthogonal-dimer chain, diagonal interaction 
terms between several particles are responsible for the series of infinite
plateaux~\cite{Schulenburg2002b}.

The effective model also includes off-diagonal processes which account for
hopping terms of bosons (i.e. triplets) from one dimer to another. In agreement with perturbation theory~\cite{Momoi2000}, 
we find that simple one-particle hopping terms are negligible. On the contrary, correlated hopping (i.e. when one particle hops from one dimer to another
 in the presence of neighboring particle) is a dominant process and we illustrate the
three most important such terms in Fig.~\ref{sh_sl_int_hop.fig} and give the
corresponding amplitudes in Fig.~\ref{sh_sl_interactions.fig}(b). 
Even though they may appear small, correlated hopping terms are very important for the physics of a system
as they favor supersolid phases~\cite{Schmidt2007a}.
We note again a good agreement with perturbation theory up to $J'/J\approx0.5$.  Beyond that, as for diagonal terms, we observe strong variations of the amplitudes 
that indicate the limit of validity of our CORE truncation. 

{\it Validity of the CORE approach --}
It is known that the physics of the 
Shastry-Sutherland model changes from a dimer state to a
2d-like Heisenberg phase above $ \big{(} J'/J\big{)}_c
\simeq 0.70$, possibly with an intermediate plaquette phase~\cite{Miyahara2003a}.
 Naturally, the question of the validity of our CORE approach emerges. 
 In this regime, we observe that (i) basic processes amplitudes do not
 converge with different cluster sizes; (ii) ground-state has zero
 overlap with our subspace. Therefore, CORE procedure is only applicable below this critical value.

Another useful tool to ascertain the validity of CORE approach is to
compute exactly on small clusters the reduced density matrix weights
of retained states~\cite{Capponi2004}. A numerical analysis done on
16-site cluster confirms that the total weight of the 2 kept states
exceeds $85\%$ as long as $J'/J\leq 0.65$. This gives us confidence
that effective interactions should decay fast enough so that our CORE
procedure is accurate in this region. Because of the reduced accuracy
close to this $J'/J$ value (which describes the SrCu$_2$(BO$_3$)$_2$
compound), we will restrict most of our findings to $J'/J=0.5$ where
various CORE truncations give similar models. Still, this value is
reasonably close to the experimental one and no qualitative changes
are expected in this region. In particular, the same magnetization plateaux 
should occur.

In all these effective models approaches, one must
distinguish the two steps: first, an effective Hamiltonian is derived;
then, because it is still an interacting quantum problem, one needs to
resort to an efficient technique to study it. Although it is a bosonic
model, the presence of positive off-diagonal terms prohibits quantum
Monte-Carlo. Possible alternatives are mean-field
analysis~\cite{Dorier2008} which main drawback is to overestimate the
tendency to form plateaux and also needs a careful numerical analysis
on finite clusters. Given the various small amplitudes, we prefer not
to make any further assumption and we provide \emph{exact
diagonalizations} (ED) of these effective models on finite lattices.

\begin{figure}
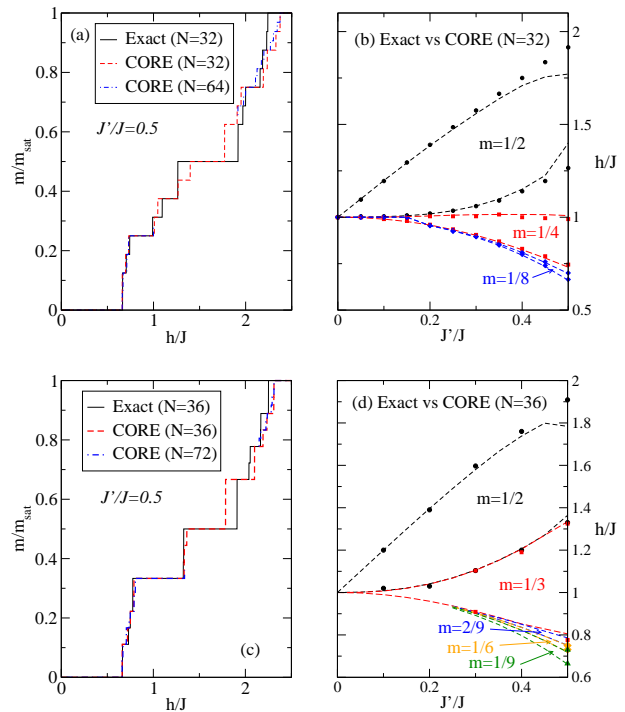

\begin{center}
\includegraphics[width=.45\textwidth,clip]{fig3a}
\end{center}
\begin{center}
\includegraphics[width=.45\textwidth,clip]{fig3b}
\end{center}
\caption[]{(Color online) (a) Magnetization curve obtained with microscopic and 
CORE models on $N=32$ lattice and CORE results on 64-site for $J'/J=0.5$.
(b) Phase diagram for $N=32$ as a function of $J'/J$ and magnetic field $h/J$: CORE results (lines) agree with ED (symbols) for
 locations of $m=1/2$, $1/4$ and $1/8$ plateaux which are allowed on this cluster. 
(c)-(d) Same as (a)-(b) for $N=36$ and 72. On these clusters,  $m=1/2$, $1/3$, $1/6$, $1/9$ or $2/9$ plateaux are 
allowed.
}
\label{Full.fig}
\end{figure}

{\it Simulations of the effective model -- } By solving exactly the effective
models for various bosonic fillings, it is straightforward to
construct the magnetization curve by a Legendre transform. A typical
plot is given in Fig.~\ref{Full.fig}(a) where we compare data
obtained with the microscopic and effective models on $N=32$ lattice. 
Because a given finite lattice cannot accommodate all magnetization
values (only multiples of $2/N$ are allowed), the magnetization curve
presents many steps. Thus, we cannot conclude yet about the
existence of other plateaux, such as 1/6, and one needs to do a
careful finite-size extrapolation to get information on the
thermodynamical limit.

Nevertheless, for this system size, Fig.~\ref{Full.fig}(b) shows that
our effective Hamiltonian is extremely accurate, as it remarkably
coincides with exact values (both for plateaux widths and locations) at
least for magnetization $m \leq 0.5$. Moreover, this accuracy is
excellent up to $J'/J\sim 0.4$ and we only observe a small disagreement for
$m=1/2$ and $J'/J=0.5$. A huge advantage of our effective model is
that, although it involves many terms (basically, all terms up to
9-body), due to Hilbert space reduction,
we are able to solve systems twice as large as for the
microscopic model, which makes feasible a finite-size scaling. We have
therefore solved ${\cal H}_{\rm eff}$ on a $N=64$ lattice and a
its magnetization curve is given in Fig.~\ref{Full.fig}(a). 
Naturally, there are twice as many finite-size steps but the main
message is that $m=1/4$ and $m=1/8$ plateaux \emph{do not change},
both in sizes and locations. This strongly suggests that our model does
exhibit magnetization plateaux in the thermodynamical limit. Because
we have chosen particular clusters, there is also the possibility to
have more stable plateaux close to these fractions, and for instance, some plateaux that we find could be unstable towards phase separation. 

Other possible fractions can be investigated by performing similar calculations 
  with other square clusters, like $N=36$ and 72, that can also 
accommodate $m=1/9$, $1/6$, $2/9$ and $1/3$. Data are shown in Fig.~\ref{Full.fig}(c-d) and confirm (i) the accuracy of our CORE effective model and
(ii) the stability of some plateaux when the system size is doubled, strongly indicating that they do persist in the thermodynamical limit. 

\begin{figure}
\begin{center}
\includegraphics[width=.4\textwidth,clip]{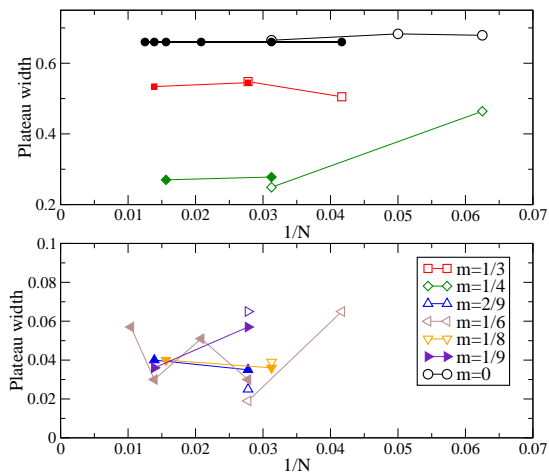}
\end{center}
\caption[]{(Color online) Finite-size scaling of the plateaux widths for various $m<1/2$ from microscopic and CORE models (respectively 
open and filled symbols) for $J'/J=0.5$ on clusters with 24, 32, 36, 48, 64, 72 and 96 sites (square and rectangular). 
Note the excellent agreement between both models solved on same lattices.} 
\label{plateau_scaling.fig}
\end{figure}

In order to give a general view of various fractions, we now turn to a
more systematic study of our effective Hamiltonian, restricted to the
typical $J'/J=0.5$ value, on various square or rectangular clusters. By computing the
magnetization curves on several square lattices, we
can perform a finite-size scaling of the plateaux widths (see
Fig.~\ref{plateau_scaling.fig}). Note that we restrict our
calculations to clusters that can accommodate a given insulating phase,
i.e. are not frustrated, according to the known
patterns~\cite{Onizuka2000,Miyahara2003a}. Since we also have access to
density-density correlations, we can also confirm these patterns (data
not shown): for instance, at $m=1/3$ (resp. $m=1/4$), the plateau is
formed by filled diagonal stripes separated by two (resp. three) empty
ones.

Our scaling provides clear evidence that there are large plateaux for $m=1/3$ and $1/4$ (besides $m=0$ and 1/2) in the thermodynamical 
limit~\cite{note}. Moreover, within the very good accuracy of our effective model, finite-size
scaling indicates smaller, but stable, plateaux for 2/9, 1/6, 1/8 and 1/9 of typical sizes between 0.01 and 0.05~$J$.

{\it Discussion --} By allowing for inhomogeneous patterns, a variety of 
fractions have been found
theoretically~\cite{Sebastian2007,Dorier2008}. Experimentally, there
is recent evidence that translation symmetry is broken for several other 
magnetization fractions~\cite{Sebastian2007,Takigawa2008}. Clearly, given the
small energy scales that stabilize one fraction with respect to
another, the accuracy of the calculation is crucial. In that sense,
systematic perturbative~\cite{Dorier2008} or CORE expansion (as in
this Letter) are promising since they are controlled techniques. For
instance, our relative error on the ground-state energy for $N=32$ and
$m=1/8$ is $0.3\%$.

However, from these effective models, a second step is to solve them
in the most unbiased way: here we have combined exact diagonalization 
and finite-size scaling and we find highly stable plateaux for 1/2, 1/3, 1/4, but also smaller plateaux for 2/9, 1/6, 1/8 and 1/9. 
While some of these fractions coincide with~\cite{Dorier2008}, we also have some differences that ask for a clarification.

A first discrepancy is the absence in our data of plateaux for $m=2/15$, observed in ~\cite{Dorier2008}, or possible other fractions such
as 1/7 or 1/5 found in~\cite{Sebastian2007}. Clearly, because our exact simulations are restricted to large but finite clusters, 
we cannot perform finite-size scaling for some fractions (as 2/15 that would require large unit cells). 
By choosing adequate shapes or boundary conditions, it might be possible to investigate some of these other fractions. 

A more crucial issue deals with $m=1/4$ and $1/8$ plateaux that are
stable in our calculations, and also found
experimentally~\cite{Kageyama1999b}, but absent in~\cite{Dorier2008}. 
A possible explanation could be that these fractions are unstable
towards phase separation but, on our finite clusters, this phenomenon
cannot be observed. However, these plateaux widths are almost constant
when the system size is doubled (see Fig.~\ref{plateau_scaling.fig})
so that we do strongly believe that they persist in the thermodynamical
limit. At this point, let us recall that our effective model is
different from~\cite{Dorier2008}, in particular for interactions beyond 2-body: for instance, 
if we only consider 2-body diagonal terms, then the $m=1/4$ plateau becomes unstable, in agreement 
with~\cite{Dorier2008}. Therefore, we think that many-body terms
ask for a careful analysis.

As a conclusion, we are convinced that reliable effective hamiltonians are crucial to understand the very rich low-magnetization properties
of the Shastry-Sutherland system. With the CORE technique, we have derived  such an effective model and, 
with finite-size scaling analysis, we provide a microscopic origin of the experimentally observed $1/3$, 1/4 and 1/8 plateaux, but we also confirm the possibility of 
other plateaux at 1/9, 1/6 and 2/9, as found in recent related studies~\cite{Sebastian2007,Takigawa2008,Dorier2008}. The remaining discrepancies 
between these approaches call for a systematic unbiased study, that would combine mean-field ideas with exact diagonalizations.

We thank  C.~Berthier, J.~Dorier, F.~Mila and K.~P.~Schmidt for
fruitful discussions and sending their results prior to publication. CALMIP (Toulouse) and IDRIS (Paris)
are acknowledged for allocation of computer time.


\end{document}